\documentclass[aps, preprint, prb]{revtex4-2}
\usepackage{graphicx, amssymb}
\usepackage{amsmath}
\usepackage{float}

\begin{document}   
   
\title{Epitaxial Stabilization and Emergent Charge Order in Copper Selenide Thin Films} 
\author{Becker Sharif}
\author{Benjamin Theunissen}
\author{Connor Westbrook}
\author{Swosti Choudhuri}
\author{Ryan Baumbach}
\author{David Lederman}
\affiliation{Department of Physics, University of California, Santa Cruz, Santa Cruz, CA 95064, USA}
\affiliation{Materials Science and Engineering Program, University of California, Santa Cruz, Santa Cruz, CA 95064, USA}
\author{Thomas Steele}
\author{Sergey Savrasov}
\affiliation{Department of Physics and Astronomy, University of California, Davis, Davis, CA 95616, USA}
\date{\today}   


\begin{abstract}
We demonstrate epitaxial growth of copper selenide (Cu$_{2-x}$Se) thin films in both cubic and rhombohedral phases, achieved via molecular beam epitaxy on Al$_2$O$3$ (001) substrates. Remarkably, the high-temperature cubic phase—which in bulk transforms into the rhombohedral structure below 400~K—is stabilized at room temperature and below, well outside its bulk equilibrium stability range. 
In the cubic phase films, temperature-dependent electrical transport reveals a pronounced, hysteretic resistivity peak near 140~K, accompanied by unit cell doubling along the [111] direction, as observed by x-ray diffraction, which are hallmarks of a charge density wave (CDW) transition. First-principles calculations show strong Fermi surface nesting in the cubic phase, consistent with the observed CDW instability. In contrast, the rhombohedral films exhibit suppressed nesting and no structural modulation. These results not only unambiguously identify a previously unreported CDW in Cu$_{2-x}$Se thin films, but also establish an epitaxial platform for tuning emergent electronic phases via strain and interface engineering.
\end{abstract}
\maketitle

\section{Main}
Copper selenide Cu$_{2-x}$Se has been the focus of extensive research due to its remarkable structural and electronic properties. In particular, it is known to exhibit a low temperature rhombohedral $\beta$-phase and a high temperature cubic $\alpha$-phase, with a range of applications in thermoelectric, ionic conduction, and emerging electronics \cite{Eikeland:lc5071,D1RA04626H}. Although the $\alpha$- and $\beta$-phases of copper selenide have been studied in their bulk forms, epitaxial thin films of these phases have not been grown to the best of our knowledge to date. Although prior studies have focused primarily on bulk thermoelectric properties, the electronic transport in thin films remains largely unexplored. 

Charge density waves (CDWs) are an example of translational symmetry breaking in the underlying crystal lattice of solid-state systems. They are characterized by periodic modulations in the electronic charge density, often coupled with lattice distortions. These phenomena predominantly arise in low-dimensional systems with favorable Fermi surface nesting conditions, where the wave vector connecting regions of high electronic density facilitates strong electron-phonon coupling~\cite{Cho2018}. CDWs have been observed in 2D~\cite{PhysRevB.14.4321}, and 3D~\cite{PhysRevLett.109.237008} systems, where their formation may not be solely driven by lattice distortions. In such cases, electron-electron interactions are presumed to play a significant role in the emergence of CDWs. Theoretical studies have shown that CDWs may compete with superconductivity, as both CDWs and superconductivity depend on effective attractive interactions---often mediated by phonons---it is worth noting that CDWs can also arise from purely electronic mechanisms, particularly when long-range interactions dominate (e.g., $U < V$ in extended Hubbard models).
 \cite{Balseiro1980}. Unlike superconductivity, which is driven by electron-electron pairing into Cooper pairs, CDWs stem from electron-hole coupling, resulting in a redistribution of charge that opens a gap in the electronic spectrum.

Experimentally, CDWs have been studied extensively in transition metal dichalcogenides (TMDs) such as TiSe$_{2}$ and TaS$_{2}$, where their formation is often accompanied by sharp anomalies in resistivity and other transport properties \cite{DiSalvo1971,Wilson1969}. These materials highlight the interplay between lattice structure and electronic behavior, providing a framework to explore similar transitions in Cu$_{2-x}$Se.


Here we report on the successful epitaxial growth of the high-temperature $\alpha$- and low-temperature $\beta$-phases of Cu$_{2-x}$Se. Remarkably, epitaxial stabilization enables the $\alpha$-phase, which is ordinarily stable only at elevated temperatures, to persist down to room temperature and below. This extended temperature range opens new opportunities to explore the intrinsic properties of the $\alpha$-phase that are inaccessible in the bulk form.  
The transport properties of the Cu$_{2-x}$Se thin films reveal signatures suggestive of CDW transitions. Such transitions are characterized by resistivity anomalies as a function of temperature, typically indicative of electron-phonon coupling leading to charge redistribution. The $\alpha$-phase thin films, in particular, exhibit behavior consistent with Fermi surface nesting, a critical condition for CDW formation.



\subsection{Crystallographic Characterization}

\begin{figure*}  
    \centering
    \includegraphics{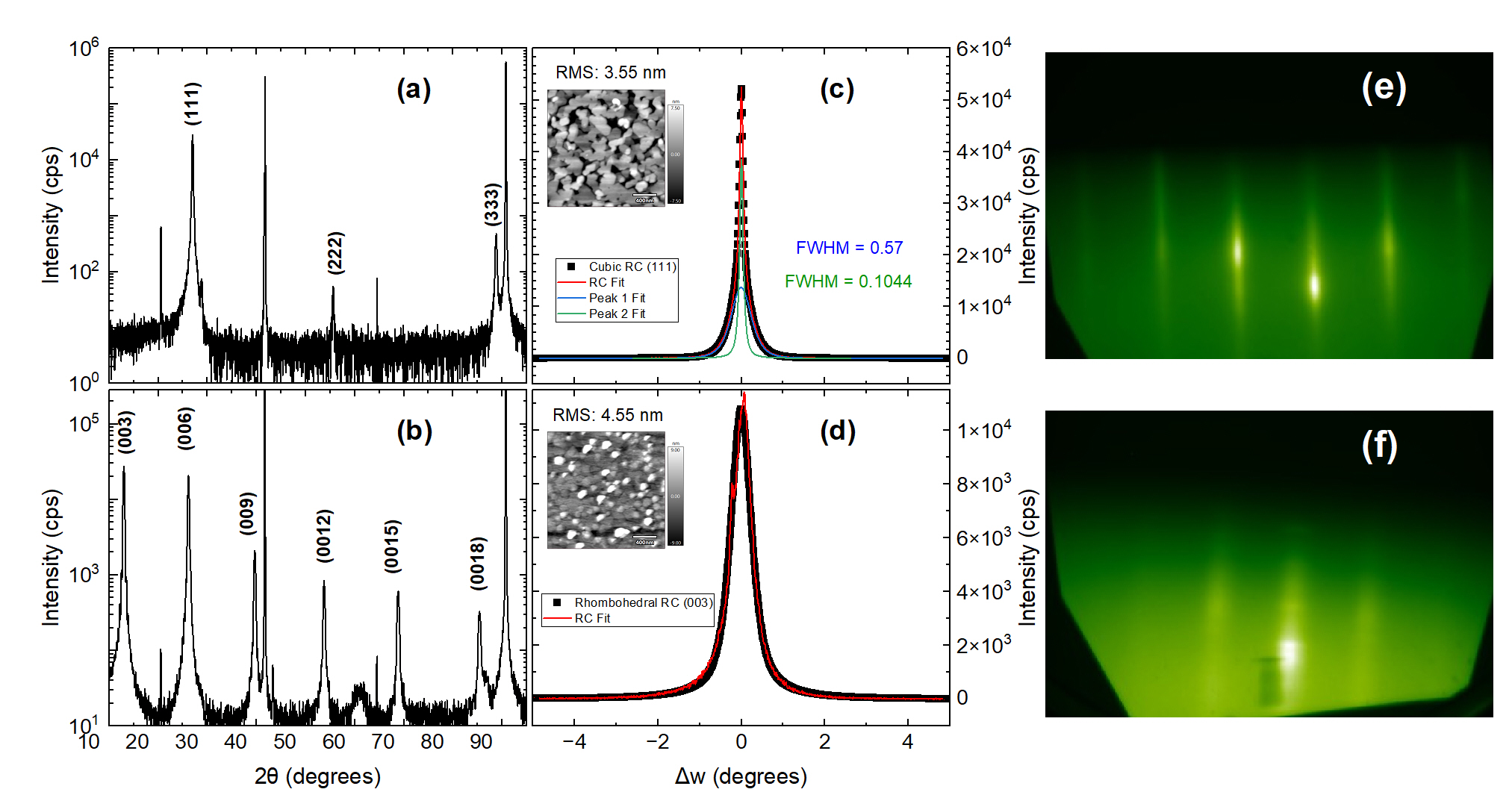}
    \caption{Comparison of structural characterization for the cubic and rhombohedral phases. (a) and (b) show XRD patterns with the scattering vector aligned along the growth direction for the cubic and rhombohedral phases, respectively. (c) and (d) display the corresponding rocking curves with AFM images shown as insets. (e) and (f) present RHEED patterns for the cubic and rhombohedral phases, respectively.
    \label{fig:RHEED_XRD_AFM}}
\end{figure*}

Room-temperature x-ray diffraction (XRD) data  for the $\alpha$-phase films, grown with high selenium flux rates, revealed a non-stoichiometric $\alpha$-phase structure of Cu$_{2-x}$Se ($0.05 \leq x\leq 0.4$) with a (111) out-of-plane orientation. The stoichiometry was found to depend on the selenium flux along with the substrate temperature. The calculated out-of-plane lattice constants ranged from $d_{111} = {3.291}$~\AA\ to $d_{111} = 3.314$~\AA, depending on the stoichiometry. Figure~\ref{fig:RHEED_XRD_AFM}(a) shows the XRD pattern of a non-stoichiometric $\alpha$-phase Cu$_{1.78}$Se (111) oriented film, with a calculated lattice constant of $a = 5.724~\AA$ ($d_{111} = 3.305$~\AA) according to a fit to the Nelson and Riley function (supplementary figure S7.a)~\cite{SIfile}. These results are in agreement with the findings of Danilkin et al.\, who reported that Cu$_{1.78}$Se belongs to the space group $Fm\bar{3}m$ with a cubic lattice parameter of $a = 5.7464(1)$~\AA~\cite{DANILKIN200357}. The thickness of the $\alpha$-phase sample was found to be 30.053~nm, according to the X-ray reflectivity (XRR) fit (supplementary figure S6.a)~\cite{SIfile}.

Rocking curve scans, shown in Fig.~\ref{fig:RHEED_XRD_AFM}(c), were performed around the (111) reflection of Cu$_{1.78}$Se. The data were fit to two peaks with full width at half maximum (FWHM) values of 0.1$^\circ$ and 0.6$^\circ$. This suggests that part of the film exhibits a high degree of out-of-plane crystalline order. Atomic force microscopy (AFM) images [inset, Fig.~\ref{fig:RHEED_XRD_AFM}(c)] revealed that the film grew in island-like structures with a typical island lateral size of approximately 0.5 $\mu$m. The island surfaces were atomically smooth, with a root mean square (RMS) roughness of 0.1~nm.

Room-temperature XRD data for the $\beta$-phase films, grown with low Se flux rates, revealed a non-stoichiometric $\beta$-phase structure of Cu$_{2-x}$Se ($0.05\leq x \leq 0.25$) with a (003) out-of-plane orientation. The stoichiometry was found to depend on the selenium flux and growth temperature. The calculated lattice constants ranged from $c = 20.413$~\AA\ to $c=20.460$~\AA, depending on the stoichiometry and growth temperature. Figure~\ref{fig:RHEED_XRD_AFM}(b) shows the XRD pattern of a non-stoichiometric $\beta$-phase Cu$_{1.95}$Se (003) oriented film, with a calculated lattice constant of $c = 20.446~\AA$ (supplementary figure S7.b)~\cite{SIfile}. These results are in agreement with the findings of Eikeland et al., who reported that Cu$_{1.95}$Se belongs to the space group $R\bar{3}m:H$ with a rhombohedral lattice parameter of $c = 20.449(1)$~\AA~\cite{Eikeland:lc5071}. The thickness of the $\beta$-phase sample was found to be 30.74~nm, according to the XRR fit (supplementary figure S6.b)~\cite{SIfile}.

Rocking curve scans, shown in Fig.~\ref{fig:RHEED_XRD_AFM}(d), were performed around the (003) reflection of $\beta$-phase Cu$_{1.95}$Se. The FWHM values were measured and fit to a single peak with an FWHM of $0.6^\circ$, which is broader than that of the $\alpha$-phase. Nevertheless, it still indicates that part of the film exhibits a high degree of out-of-plane and in-plane crystalline order. Atomic force microscopy (AFM) images for the $\beta$-phase Cu$_{1.95}$Se films [inset, Fig.~\ref{fig:RHEED_XRD_AFM}(d)] revealed that the film grew in island-like structures with a typical island size of approximately 0.5~$\mu$m. The island surfaces were atomically smooth, with an RMS roughness of 0.1~nm.

Reflection high-energy electron diffraction (RHEED) patterns for both phases confirmed the epitaxy of the films. Figures~\ref{fig:RHEED_XRD_AFM}(e) and~\ref{fig:RHEED_XRD_AFM}(f) show representative streaky RHEED patterns for the cubic and rhombohedral phases, respectively, observed during and after film growth. The bright, sharp streaks are indicative of smooth, crystalline surfaces (RHEED, AFM and XRD data for a second $\alpha$- and a $\beta$-phase sample are shown in supplementary figure S8~\cite{SIfile} ).

\begin{figure*}  
   \includegraphics{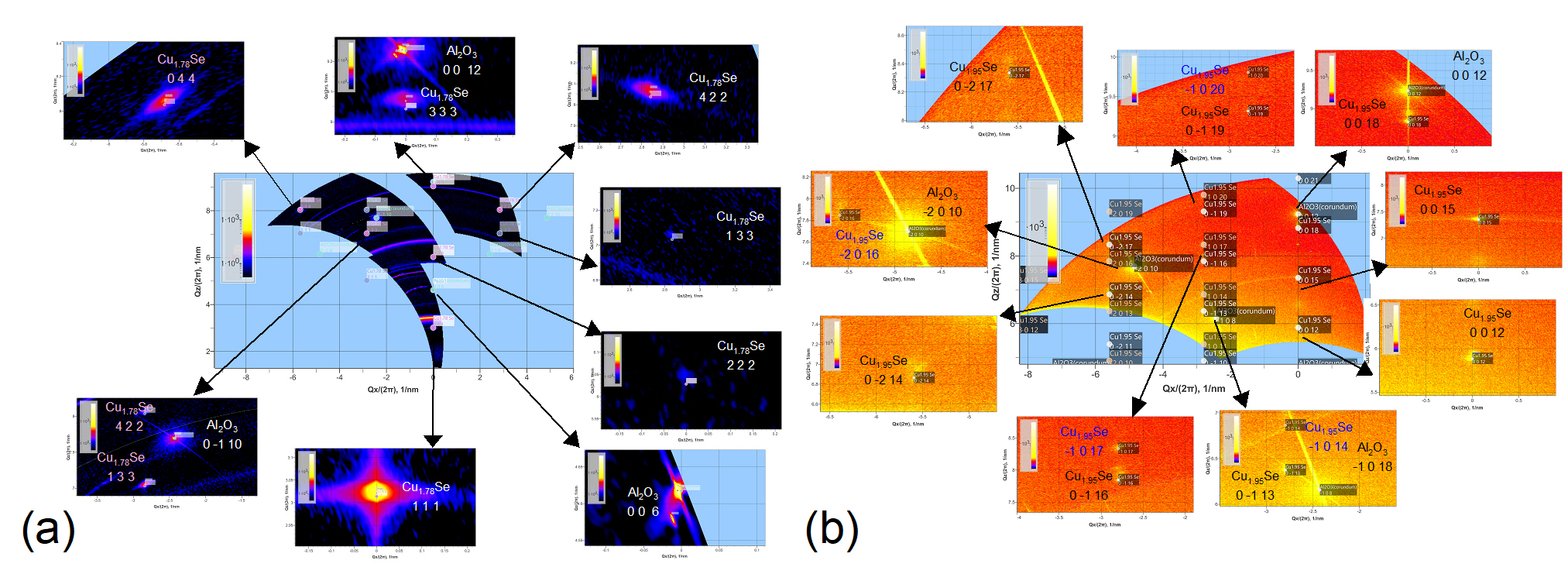}
    \caption{Reciprocal space maps (RSM) of (a) the cubic phase sample and (b) the rhombohedral phase sample. The different color indices correspond to distinct in-plane twinned domains, with each twin orientation represented by a separate color. 
    \label{fig:RSM}}
\end{figure*}

The strain of the films was further analyzed using Reciprocal Space Mapping (RSM). Figures ~\ref{fig:RSM}(a) and (b) show the RSM data for the cubic and rhombohedral phases, respectively. In the images, we observe peaks with pure out-of-plane components and peaks with both in-plane and out-of-plane components. For the cubic phase, we detect the (111) peak along with its higher-order reflections, while for the rhombohedral phase, we observe the (003) peak and its higher-order reflections. Additionally, the presence of the (422) peak in the cubic phase and the $(0\overline{1}13)$
 peak in the rhombohedral phase confirms contributions from both in-plane and out-of-plane lattice components.

From these peaks, we calculated the strain in the films. The calculations, detailed in supplementary sections 1 and 2~\cite{SIfile}, show that the cubic phase exhibits an in-plane lattice contraction of approximately $0.07\%$ , indicating that the cubic samples are not strained relative to their bulk counterparts. Similarly, in the rhombohedral phase, we observe a $0.04\%$ in-plane contraction,  indicating that the rhombohedral samples are not strained relative to their bulk counterparts.

Additionally, RSM measurements reveal 60$^\circ$ in-plane twinning in both phases. In the cubic phase, this is evident by the appearance of the (422) and (133) reflections at two distinct in-plane azimuthal positions, separated by 60$^\circ$. Given the (111) out-of-plane orientation of the film—which results in a hexagonal in-plane lattice—this repeated symmetry is consistent with rotational twinning among equivalent $\langle 111 \rangle$ domains. As for the rhombohedral phase, the $(0\overline{1}13)$ reflection appears alongside the $(\overline{1}014)$ reflection. The presence of the $(\overline{1}014)$ reflection at this specific position can only be explained by a distinct in-plane azimuthal rotation of 60$^\circ$. Given the (003) out-of-plane orientation of the films, with a hexagonal in-plane lattice, this repeated symetry is also consistent with rotational twinning among equivalent $\langle 003 \rangle$ domains. Both the cubic and rhombohedral phases exhibit this twinning behavior.

\subsection{Electronic Transport Measurements}

\begin{figure}
    \centering
    \includegraphics[width=\linewidth]{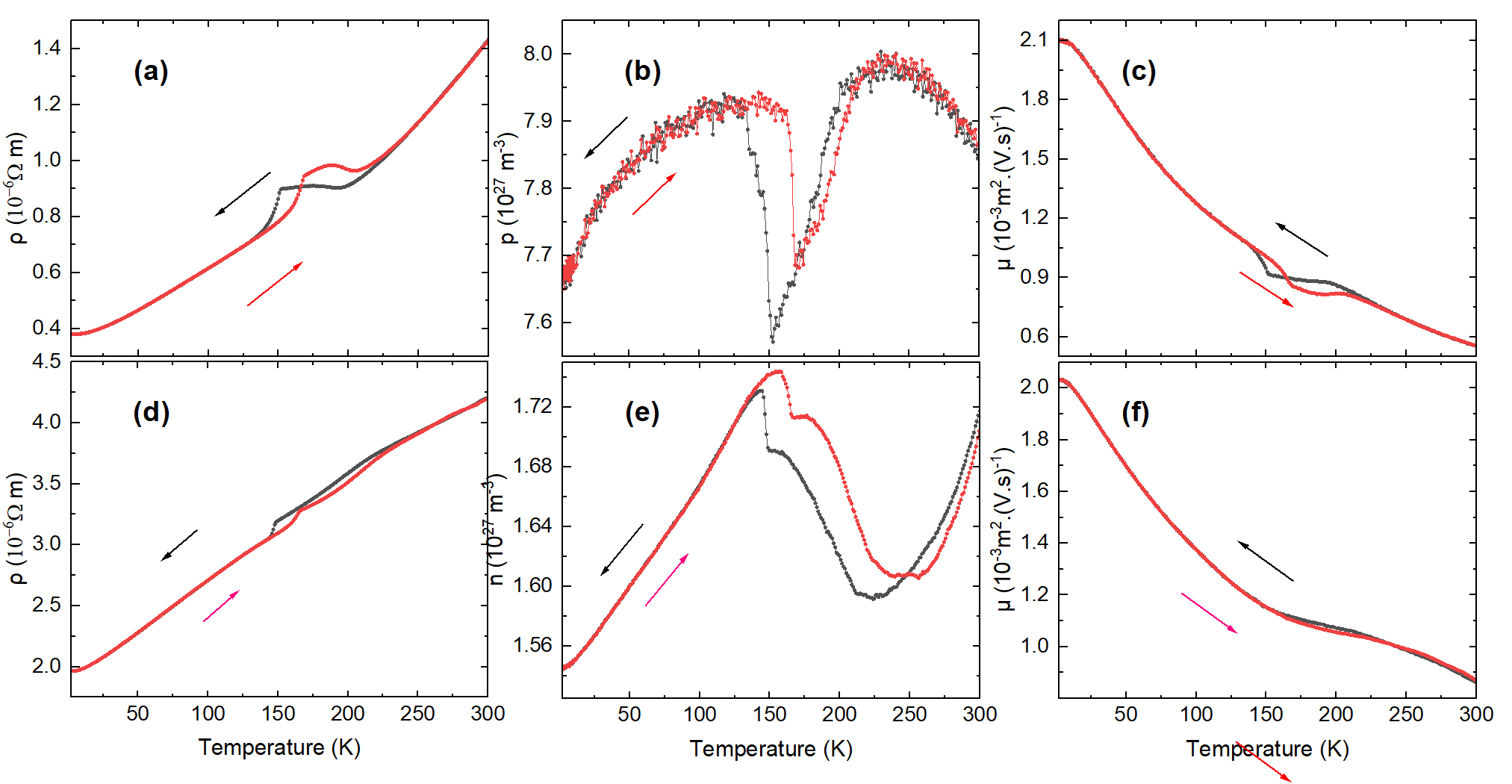}
    \caption{Resisitivity, carrier density, and mobility as functions of temperature for the cubic [(a)-(c)] and the rhombohedral [(d)-(f)] samples.
    \label{fig:transport}}
\end{figure}

Figure~\ref{fig:transport} show the resistivity ($\rho$) carrier concentration ($p$) and mobility ($\mu$) as functions of temperature measurements for the $\alpha$-phase (cubic) and $\beta$-phase (rhombohedral) Cu$_{2-x}$Se thin films. A clear metal-to-insulator transition is observed around 190 K during cooling and around 200 K during warming in the cubic phase, indicating hysteresis. Additionally, an insulator-to-metal transition is observed around 150 K during cooling and around 165 K during warming. The repeatability of these transitions across multiple measurements suggests an intrinsic electronic ordering phenomenon (resistivity ($\rho$) carrier concentration ($p$) and mobility ($\mu$) as functions of temperature data for a second $\alpha$- and a $\beta$-phase sample are shown in supplementary figure S10~\cite{SIfile} ).

We propose that these transitions correspond to a charge density wave (CDW) phase transition, similar to what has been observed in materials like TiSe$_2$ and TaS$_2$, where resistivity anomalies occur due to electron-phonon coupling and the formation of a periodic charge modulation \cite{DiSalvo1971, Wilson1969, PhysRevB.14.4321}. The hysteresis in these transitions, as depicted in Figure ~\ref{fig:transport}(a), is characteristic of a first-order phase transition. Similar hysteretic behavior has been reported in other CDW systems, such as 1T-TaS$2$, where the transition is driven by lattice distortion and electronic instability \cite{Cho2018}. This observation further supports the hypothesis that the phase transitions in $\alpha$-phase Cu$_{2-x}$Se are linked to a CDW instability.

When compared to the rhombohedral phase, Fig.~\ref{fig:transport}(b) shows that this phase exhibits a suppression of this phase transition. The absence of a CDW transition in the $\beta$-phase suggests that the structural configuration of the rhombohedral phase suppresses the instability observed in the cubic phase. The relationship between crystal structure and CDW formation is well-documented in various materials, where symmetry and dimensionality play critical roles in stabilizing or inhibiting CDW behavior \cite{Balseiro1980, PhysRevLett.109.237008}.

\subsection{XRD vs. Temperature Measurements}
\begin{figure}
    \centering
    \includegraphics[width=\linewidth]{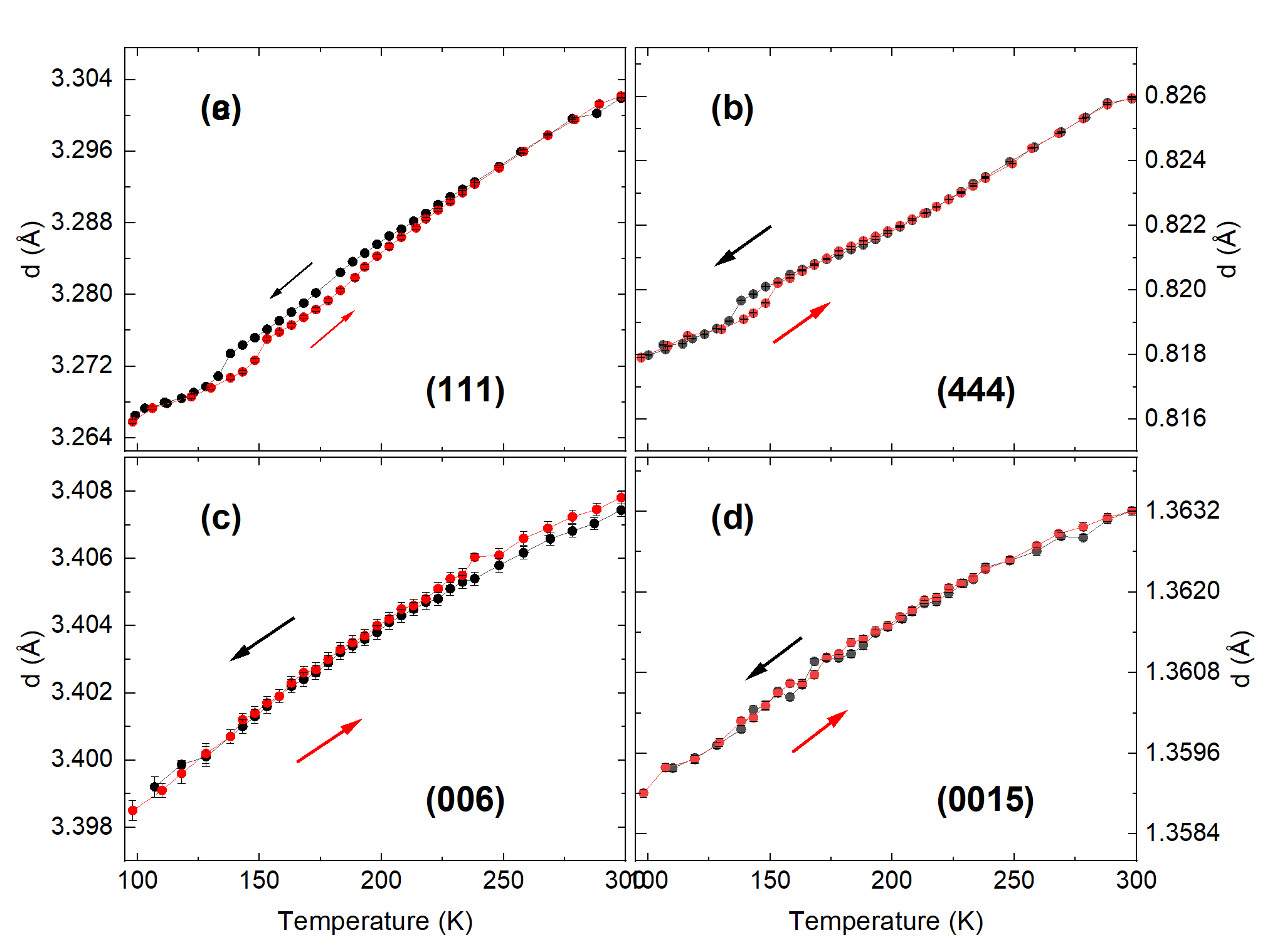}
    \caption{XRD vs. temperature: Comparison of $d$-spacing vs. temperature for the cubic phase—(a) (111), (b) (444) —and the rhombohedral phase—(c) (003), (d) (00 15) peaks.
    \label{fig:XRD vs T}}
\end{figure}

XRD d-spacing measurements as a function of temperature for the (111) and (444) peaks in the $\alpha$-Cu$_{2-x}$Se (cubic-phase) thin films—shown in Figures ~\ref{fig:XRD vs T}(a) and (b), respectively, (supplementary figure S3.c shows the (422) peak with both in-plane and out-of-plane components)~\cite{SIfile}—reveal an anomalous change in slope at $T=187$~K during cooling and at $T=197$~K during warming. This anomaly is consistent with the resistivity transitions observed in the transport measurements. Notably, the (111) and (444) peaks have only out-of-plane components, while the (422) peak contains both in-plane and out-of-plane components. The anomalous behavior in d-spacing is also accompanied by changes in peak intensity and full-width at half maximum (FWHM), as shown in supplementary figure S4~\cite{SIfile}.

These structural anomalies suggest a coupling between lattice distortions and electronic instabilities, further supporting the possibility of a CDW transition. CDW transitions are often associated with periodic lattice distortions that manifest as shifts in peak positions, changes in peak intensities, and broadening of diffraction peaks \cite{Balseiro1980}. Similar structural behavior has been reported in TiSe$_2$, where the onset of the CDW phase correlates with a softening of phonon modes and a redistribution of electron density~\cite{DiSalvo1971, Wilson1969}.

\begin{figure}
        \includegraphics[width=0.5\linewidth]{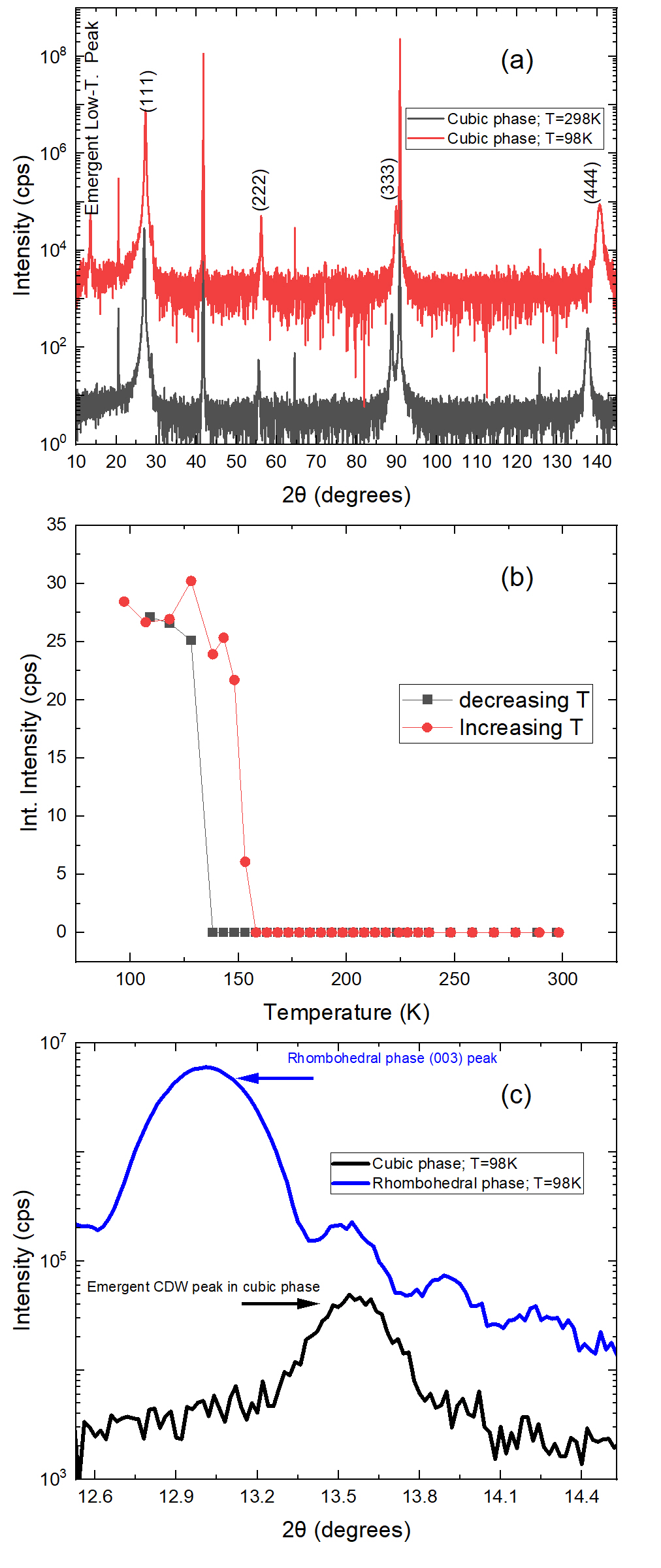}
    \caption{(a)XRD $\theta-2\theta$ scans of the cubic phase sample at $T=98$~K (red) and $T=298$~K (black) showing the emergent low-temperature peak. (b) Integrated intensity of the emergent CDW peak for the cubic sample as a function of temperature. (c) $\theta-2\theta$ scan in the low angle region at $T=98$~K for the rhombohedral phase sample's (003) peak (blue) and the emergent CDW peak of the cubic phase sample (black). Notice the significant difference in peak position, indicating that the emergent peak is not a transition to the rhombohedral phase.
    \label{fig:CuSe092124 and CuSe091324 emergent peak analysis}}
\end{figure}

In addition to these $d$-spacing anomalies, we observe the emergence of a super-lattice peak at temperatures below the phase transition identified in transport measurements. Figure~\ref{fig:CuSe092124 and CuSe091324 emergent peak analysis}.a shows this emergent peak at $T=98$~K. Figure ~\ref{fig:CuSe092124 and CuSe091324 emergent peak analysis}.b presents the intensity of this emergent superlattice peak as a function of temperature. The superlattice peak appears at $T=187$~K during cooling and disappears at $T=197$~K during warming, in agreement with both the resistivity and $d$-spacing anomalies. Figure~\ref{fig:CuSe092124 and CuSe091324 emergent peak analysis}.c shows the $\theta-2\theta$ scan for both the cubic and rhombohedral phase samples at $T=98$~K. In this comparison, we focus on the (003) peak of the rhombohedral phase and the emergent superlattice peak observed in the cubic phase. The figure clearly demonstrates that these peaks occur at different positions, confirming that the emergent peak in the cubic phase is not due to a structural phase transition. Notably, the emergent peak appears at exactly half the (111) Bragg peak position, indicating a doubling of the lattice periodicity. This suggests that the observed peak is a CDW superlattice reflection with a modulation period twice that of the underlying lattice.

In contrast, XRD measurements of the (006) and (0015) peaks in the $\beta$-Cu$_{2-x}$Se (rhombohedral-phase) thin films, shown in Figures ~\ref{fig:XRD vs T}(c) and (d), respectively (supplementary figure S5 also shows the (003) (0018) peaks)~\cite{SIfile} exhibit no similar anomalous behavior. Unlike the cubic phase, the rhombohedral phase does not show a clear transition in resistivity, and no superlattice reflections are detected. The absence of these signatures suggests that the CDW instability observed in the cubic phase is suppressed in the rhombohedral phase, reinforcing the importance of the role of crystal symmetry and electronic structure in stabilizing or inhibiting charge density wave formation.

\subsection{DFT Calculations}

\begin{figure}[H]
    \centering
    \includegraphics[width=\textwidth]{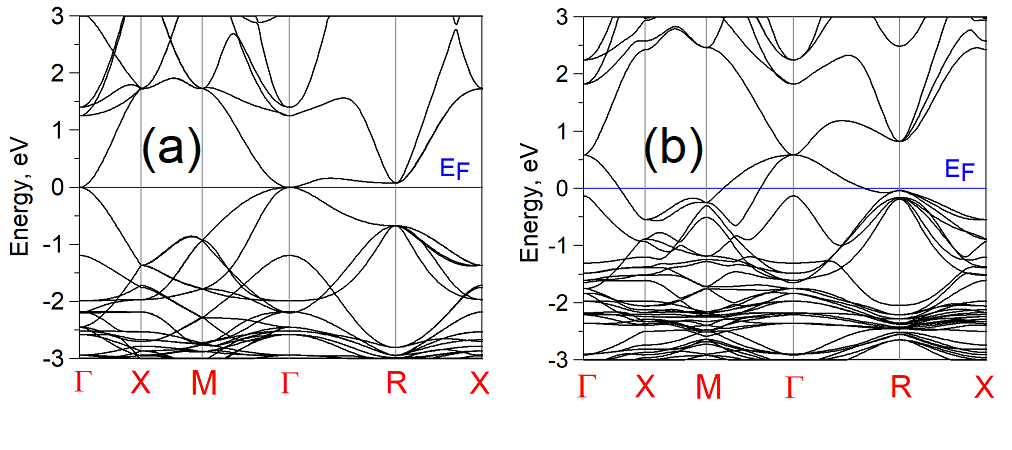}%
    \caption{Calculated band structure of (left) cubic Cu$_{2}$Se and (right) cubic Cu$_{1.75}$Se using
the unit cell of primitive cubic lattice with 4 formula units. The dispersions are shown
along high-symmetry lines of the Brillouin Zone of the cubic lattice (see Figure ~\ref{fig:Fermi Surface - cubic} for the
notations)
    \label{fig:Cubic Bands}}
\end{figure}

\begin{figure}
    \centering
    \includegraphics[width=0.4\linewidth]{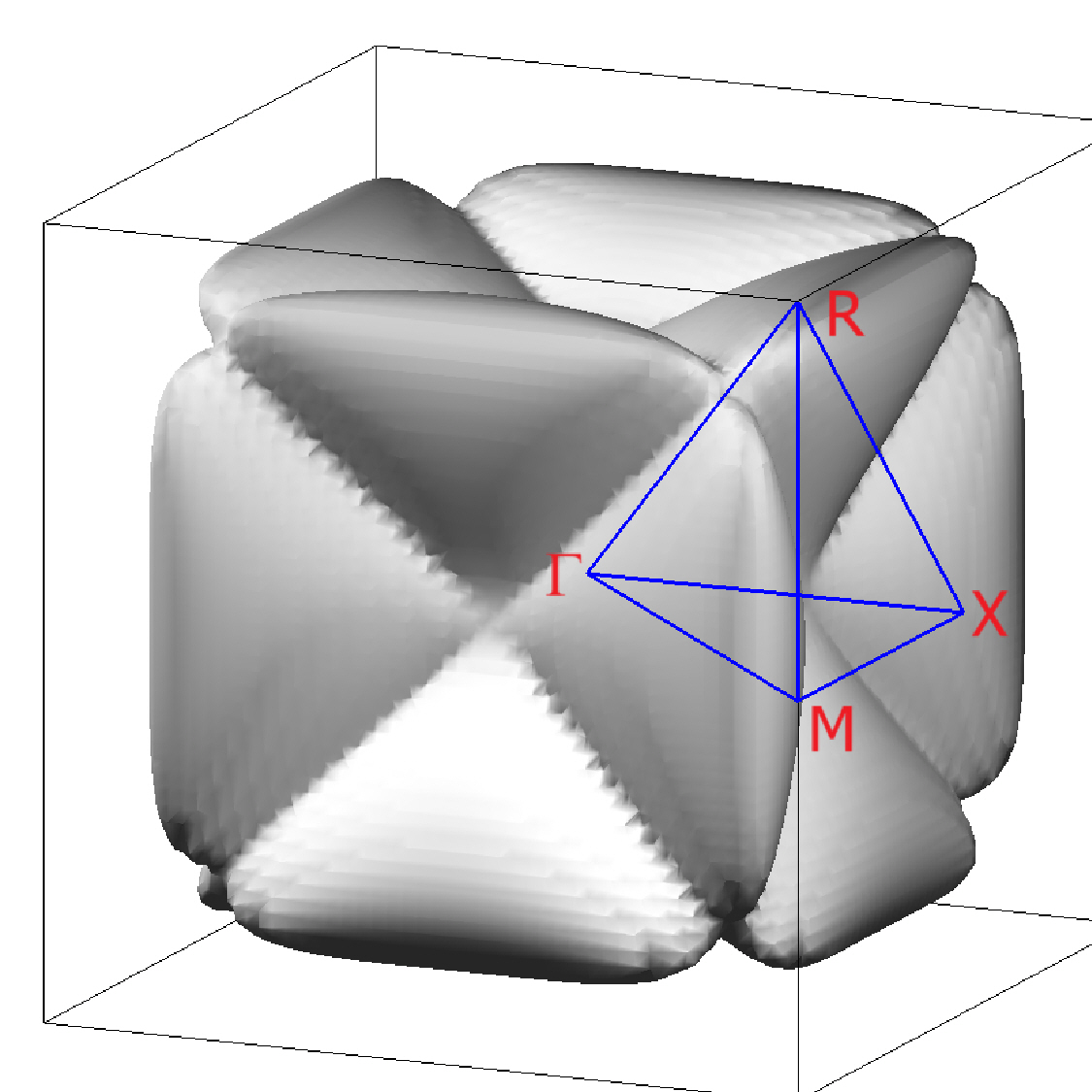}
    \caption{Calculated Fermi surface of cubic Cu$_{1.75}$Se using the unit cell of primitive
cubic lattice with 4 formula units (Cu$_{7}$Se$_{4}$). The Fermi surface is shown in the cubic
Brillouin Zone whose irreducible part with high-symmetry points is plotted. Flat areas of
this Fermi surface indicate strong nesting features corresponding to the wavevector
(1/2, 0, 0).
    \label{fig:Fermi Surface - cubic}}
\end{figure}

Figures~\ref{fig:Cubic Bands}(a) and (b) show the band structure calculation for stoichiometric $\alpha$-Cu$_{2}$Se and non stoichiometric $\alpha$-Cu$_{1.75}$Se, respectively. Both calculations are done in the unit cell
of primitive cubic structure, and the results for the band dispersion are shown along high
symmetry lines of the Brillouin Zone (BZ) of the cubic lattice (see notations for high symmetry
points in Fig.~\ref{fig:Fermi Surface - cubic}. The calculations show that stoichiometric $\alpha$-Cu$_{2}$Se is a zero-gap
semiconductor. As the Cu concentration decreases, the Fermi level shifts down and we
expect the pocket around the $\Gamma$ point of the Brillouin zone to develop. The
calculations show that the non-stoichiometric $\alpha$-Cu$_{2-x}$Se is a hole doped
semiconductor, in agreement with our transport measurements. We have studied the
effect of disorder by taking out one copper atom from its various positions in the cubic
unit cell but found the results to be practically similar. A comparison of figures~\ref{fig:Cubic Bands}(a) and (b) also shows that the removal of copper can be simply modeled by the rigid band
picture.

\begin{figure}
    \centering
    \includegraphics[width=0.4\linewidth]{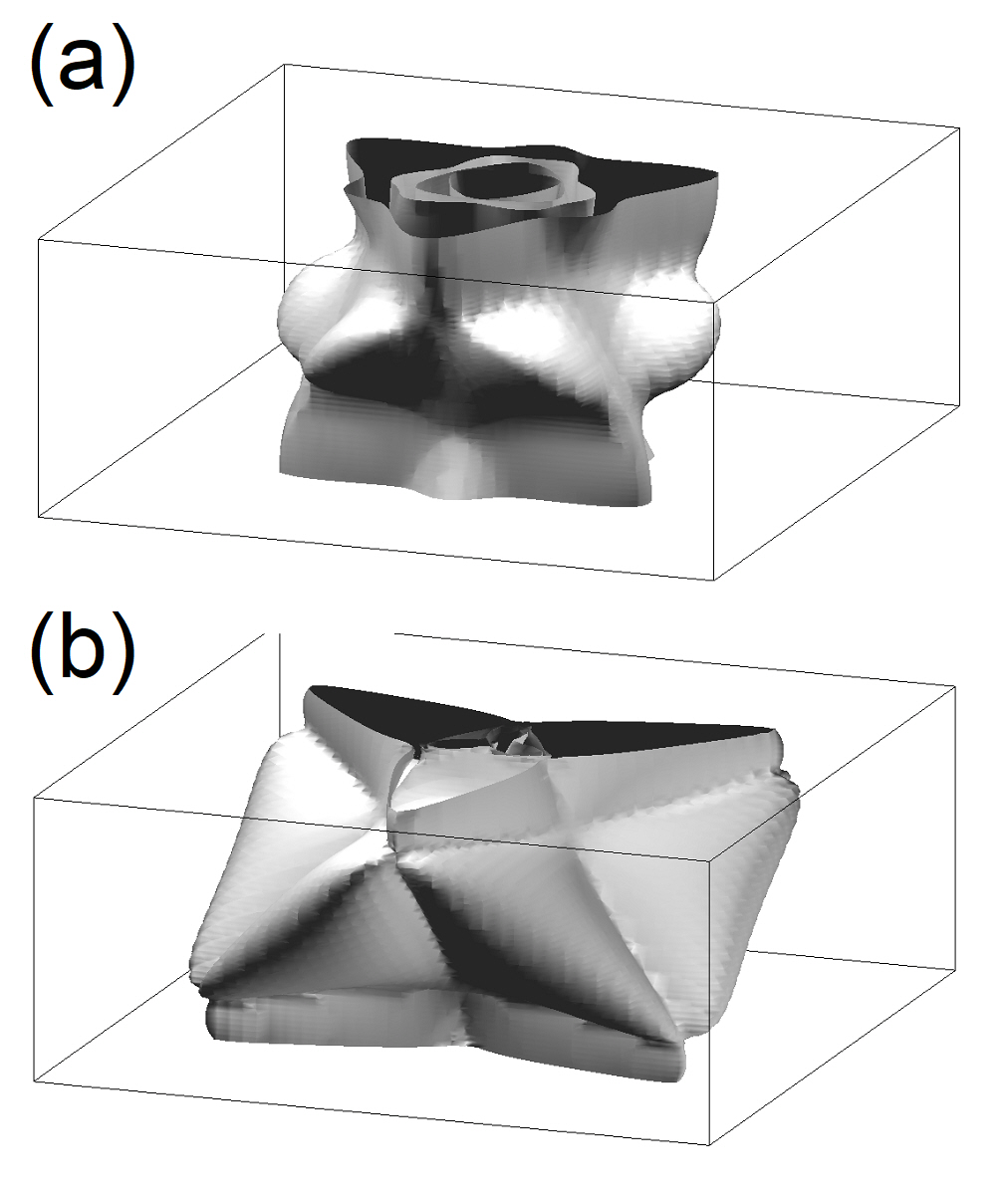}
    \caption{(a) Calculated Fermi surface of rhombohedral phase of Cu$_{1.75}$Se with no
clear evidence for the nesting. (b) Fermi surface of cubic Cu$_{1.75}$Se replotted in
rhombohedral coordinates to highlight the difference that rhombohedral distortions bring
in suppressing the nesting features seen in the cubic phase.
    \label{fig:Fermi Surface - rhombohedral vs cubic}}
\end{figure}

Figure~\ref{fig:Fermi Surface - cubic} shows the calculated Fermi surface of non-stoichiometric cubic Cu$_{1.75}$Se
together with the irreducible Brillouin Zone of the cubic lattice. A strong nesting feature
is observed along the [100] direction ($\Gamma$–X), with a pronounced peak at the wavevector
(1/2, 0, 0). This nesting is further supported by the calculated electronic susceptibility
function. These theoretical results are consistent with both transport and X-ray
diffraction measurements, providing strong evidence that the CDW observed in the
cubic phase is driven by Fermi surface nesting.
We have performed similar density functional studies of the rhombohedral Cu$_{1.75}$Se
using the crystal structure found in \cite{Eikeland:lc5071}. Figure ~\ref{fig:Fermi Surface - rhombohedral vs cubic}(a) shows its calculated Fermi
surface. Although the Brillouin Zone of the rhomboredral phase is different, one cannot
clearly resolve flat areas of the Fermi surface seen in Figure~\ref{fig:Fermi Surface - cubic}  of the cubic phase,
thus one can conclude that the nesting is suppressed.

To gain additional insight, we have done the calculation of the cubic phase using the
primitive unit cell of the rhombohedral lattice since the latter can be viewed as the
evolution of the cubic phase: if the z-axis is chosen along [111] direction, the cubic
phase corresponds to ABC stacking of Se atoms and Copper atoms occupying
tetrahedral sites with the $c/a = \sqrt{6} = 2.449$. The rhombohedral $\beta$-Cu$_{1.75}$Se corresponds to doubling the unit cell along [111] and a slightly stretched c/a ratio equal
to 2.503(x2) with respect to the cubic lattice.
The Fermi surface of the cubic phase in the Brillouin Zone of the rhombohedral lattice is
shown in Figure~\ref{fig:Fermi Surface - rhombohedral vs cubic}(b). It is technically equivalent to the calculation of the cubic phase
shown in Figure~\ref{fig:Fermi Surface - cubic} but redrawing it in the rhombohedral coordinates allows us a direct
comparison with the Fermi surface of $\beta$-Cu$_{1.75}$Se. One can clearly see how the nesting
is destroyed as the rhombohedral phase develops.

The presence of a CDW in the cubic phase and its suppression in the rhombohedral phase highlights the significant influence of crystal symmetry on electronic ordering. In the cubic phase, enhanced Fermi surface nesting conditions and strong electron-phonon coupling contribute to CDW formation, whereas the rhombohedral phase disrupts these nesting conditions, preventing the emergence of the instability. This is consistent with prior studies on transition metal dichalcogenides and other CDW materials, such as NbSe$_2$, where disorder and dimensionality play a crucial role in determining CDW stability \cite{Cho2018}.

Comparisons to other CDW systems further emphasize the role of structural tuning in controlling CDW emergence. In TiSe$_2$, for example, stacking sequences and subtle distortions dictate the onset of CDW order, while in (Sr,Ca)$_3$Ir$4$Sn${13}$, pressure and compositional changes can induce or suppress CDW behavior \cite{PhysRevB.14.4321, PhysRevLett.109.237008}. Similar structural effects have been observed in Cu$_2$Se, where superionic transitions and lattice fluctuations influence charge transport \cite{DANILKIN200357, YAMAMOTO1991202}.

\section{Methods}

\subsection{Thin Film Deposition}
Cu$_{2-x}$Se thin films were grown in an ultra high vacuum (UHV) molecular beam epitaxy (MBE) system (base pressure $< 10^{-9}$ Torr) by sublimation of copper and selenium pallets ($>99.999\%$ purity) onto polished Al$_2$O$_3$ (0001) substrates. Before film growth, the substrates were annealed at 1200~$^\circ$C for 2 hours in an atmospheric annealing furnace (Lindberg/Blue M), followed by a secondary annealing in UHV at 800~$^\circ$C for 2 hours. Reflection high-energy electron diffraction (RHEED) patterns were acquired after the substrate was annealed to ensure satisfactory surface smoothness and crystallinity. A retractable quartz crystal monitor inside the growth chamber was used to calibrate the molecular flux of the copper and selenium beams and to set the desired phase and stoichiometry of each sample. 

The growth process consisted of a continuous evaporation of copper at a rate of approximately 2.75~\AA/min, while selenium evaporation was alternated using a shutter programmed to open and close in cycles. The selenium shutter was closed for 78 seconds and open for 63 seconds. The substrate temperature was held at $450~^\circ\mathrm{C}$ during the growth. The $\alpha$- phase was grown under conditions of relatively high selenium flux (ranging from 10 to 40 \AA/min), where the stoichiometry of the final thin film depended on the selenium flux rate. The $\beta$- phase was highly sensitive to the selenium flux rate and required a relatively low flux rate of approximately 3.05 \AA/min for successful growth. The initial growth pressure of the chamber was approximately \(\sim 2 \times 10^{-10}\)~Torr, with the pressure rising to \(\sim 6 \times 10^{-8}\)~Torr during deposition. 

\subsection{Crystallographic Characterization}
Crystal quality was monitored during sample growth using \textit{in-situ} reflection high energy electron diffraction (RHEED) and sample topography was measured after growth using tapping mode atomic force microscopy (AFM) at room temperature (Oxford Cypher AFM). A Rigaku SmartLab x-ray diffractometer with a Cu  rotating anode and a Ge double-bounce monochromator to select K$_{\alpha_1}$ radiation (wavelength $\lambda=0.15406$~nm) was used for x-ray diffraction (XRD), x-ray reflectivity (XRR), and reciprocal space mapping (RSM) measurements after growth. Temperature-dependent XRD and RSM measurements were performed using a liquid nitrogen (LN$_2$) cryogenic stage. The XRD and RSM data were analyzed quantitatively using  symmetric pseudo-Voigt peak fitting with Rigaku's SmartLab software to determine the lattice constants and interface roughness parameters. XRR measurements were used to determine thin film thicknesses. The resulting data were analyzed using the XRR fitting module in SmartLab Studio II. A multilayer model was constructed for each sample, and the reflectivity curves were fit using a model-based approach that employs Parratt's formalism combined with global optimization algorithms to minimize the difference between the simulated and experimental data.

The value of the lattice parameter $a$ out of the plane of the sample was calculated from the XRD peak positions according to Bragg’s law, $2d_{hkl} \sin\theta_{hkl}=\lambda$ , where $d_{hkl}$ is the lattice constant corresponding to planes defined by the Miller indices $(hkl)$ and, $\theta_{hkl}$ is the measured Bragg diffraction angle corresponding to the $(hkl)$ plane. The lattice parameters for both the $\alpha$ - and $\beta$ - phases were then determined with greater experimental precision with higher-order peaks using the Nelson and Riley function
\begin{equation}
\frac{\Delta a}{a} = \kappa \left( \frac{\cos^2\theta}{\sin\theta} + \frac{\cos^2\theta}{\theta} \right).
\end{equation}
The lattice constant was determined by plotting the calculated $a$ values as a function of the Nelson-Riley function and then fitting the resulting data to a straight line. 
\subsection{Devise Fabrication and Electronic Transport Measurements}
$\alpha$- and $\beta$- phase samples were made into Hall bars for transport measurements using photolithography, followed by argon ion etching in a UHV ion chamber to remove the unwanted film. Electrode pads were then patterned using photolithography, followed by chromium-gold deposition and photoresist lift-off to remove the unwanted chromium-gold. The finished Hall bars were then adhered to chip carriers using conducting silver paint. Electrical contacts were made between the Hall bar pads and the chip carrier contacts by wedge wire bonding. The Hall bars were 250~$\mu$m wide and contacts were separated by 670~$\mu$m. The Hall bar samples were measured using two systems: a Janis 12TM-SVM Super VariTemp cryostat, and a Quantum Design PPMS. Both systems can measure at temperatures ranging from 350~K to 2~K and fields of up to 11~T and 9~T, respectively. Electronic measurements were made by sourcing 100 $\mu$A of current using an SRS CS580 voltage controlled current source and voltage measurements were taken using EG\&G Instruments 7265 and Stanford Research Systems SR830 lock-in amplifiers at 13.77 Hz.

Calculations of carrier density ($n$) were performed based on the measured Hall resistances along with film thickness using $n = 1/R_{xy}et$, where $e$ is the charge of the electron and $t$ is the film thickness. The Hall mobility $\mu$ was calculated from the measured longitudinal resistivity $\rho_{xx}$ of the device together with the carrier density $n$, using $\mu = 1/ne\rho_{xx}$. A magnetic field of 9~T was applied during the Hall measurements.
\subsection{DFT Calculations}
We perform Density Functional Calculations for both cubic and rhombohedral phases of
Cu$_{2-x}$Se using full potential linear muffin tin orbital method \cite{PhysRevB.54.16470}. The
stoichiometric cubic Cu$_{2}$Se is realized in the fluoride structure. Non-stoichiometricity
can be modelled by considering a simple cubic structure with 4 formula units (Cu$_{8}$Se)$_{4}$)
and then removing one copper atom (Cu$_{7}$Se)$_{4}$ or Cu$_{1.75}$Se)) which is close to the
experimentally grown samples.

\section{Outlook}
In this study, we investigated the structural and transport properties of epitaxial Cu$_{2-x}$Se thin films, focusing on the role of crystal structure in electronic phase transitions. Using molecular beam epitaxy, we successfully grew high-quality cubic $\alpha$- and rhombohedral $\beta$- phase films on Al$_2$O$_3$ (0001) substrates. Structural characterization via XRD, RHEED, and AFM confirmed the epitaxial quality of the films, with distinct lattice parameters and strain states for each phase.

Temperature-dependent transport measurements revealed a hysteretic phase transition in the cubic phase, characterized by resistivity anomalies at approximately 190 K and 150 K. These transitions suggest the presence of a CDW. XRD versus temperature measurements further supported this interpretation, revealing an anomalous change in lattice spacing and the emergence of a superlattice reflection at low temperatures in the cubic phase, indicative of periodic lattice distortions. In contrast, the rhombohedral phase exhibited no such electronic transition, underscoring the role of crystal symmetry in CDW formation. Additionally, DFT calculations confirmed strong Fermi surface nesting in the cubic phase, consistent with the observed structural anomalies. In contrast, the rhombohedral phase showed a suppression of Fermi surface nesting, further highlighting the symmetry-dependent nature of CDW formation.

Our results demonstrate that structural phase selection plays a crucial role in stabilizing or suppressing CDW behavior in Cu$_{2-x}$Se. The cubic phase, with its favorable Fermi surface nesting conditions, appears to support CDW formation, whereas the rhombohedral phase does not. These findings provide new insights into the interplay between structure, electronic ordering, and transport properties in copper selenide. Stabilizing other materials that potentially undergo a CDW transition could lead to new functionalities with implications for potential applications in electronic and quantum devices.

Future work could explore the impact of epitaxial strain, doping, and dimensional confinement on CDW behavior in Cu$_{2-x}$Se, further elucidating the mechanisms governing its phase transitions. Additionally, spectroscopic techniques such as angle-resolved photoemission (ARPES) and scanning tunneling microscopy (STM) could provide direct evidence of charge modulation and electronic band reconstruction, offering deeper insight into the electronic phase diagram of this material.

\clearpage
\appendix
\renewcommand{\appendixname}{Supplementary Material}
\renewcommand{\thesection}{S\arabic{section}}
\section*{Supplementary Material}
\addcontentsline{toc}{section}{Supplementary Material}
\renewcommand{\thefigure}{S\arabic{figure}}
\setcounter{figure}{0}



\title{Supplementary Materials: Epitaxial Stabilization and Emergent Charge Order in Copper
Selenide Thin Films}
\author{}

\hrulefill
\vspace{10pt}
\section{RSM and Strain - cubic phase}
\vspace{10pt}

\begin{figure}[H]  
    \centering
    \includegraphics[width=\linewidth]{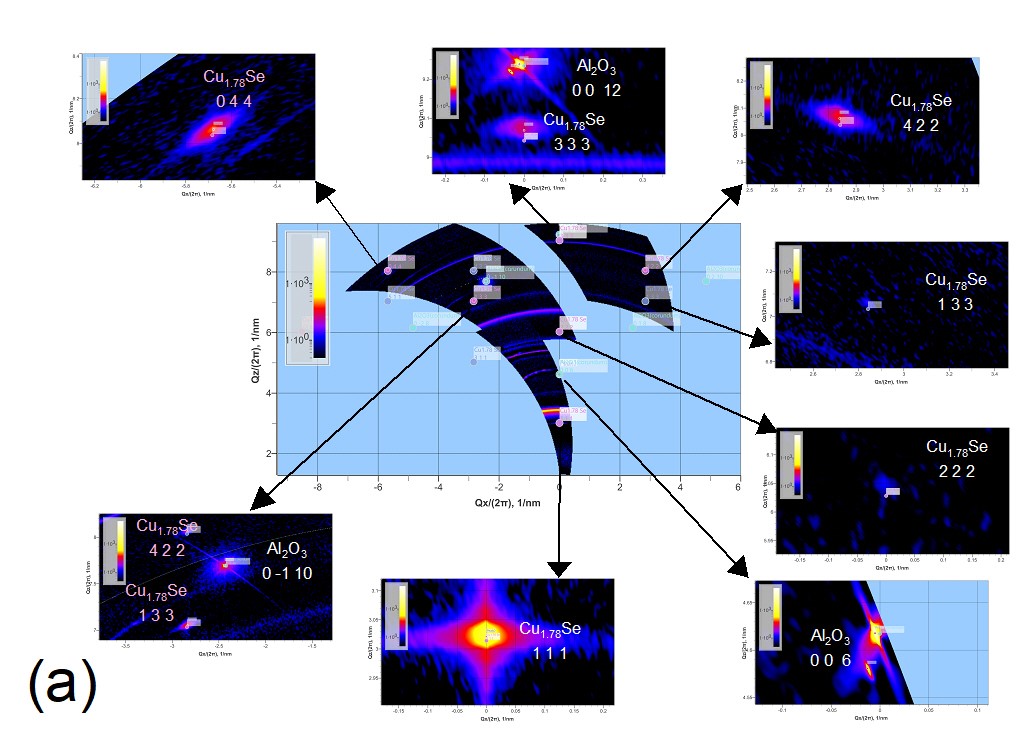}
    \caption{Detailed RSM comparison of Cubic phase.}
    \label{fig:cubic_phase}

\end{figure}

\textbf{Figure 1 shows the RSM for a cubic phase sample. We will focuse on the (4 2 2) peak during this calculation and compare the results with Cu$_{1.78}$Se from COD ID 1532676. FROM RSM scan we have:}

\[
\frac{Q_{x(422)}}{2\pi} = -2.8439 \pm 0.00004~\text{nm}^{-1}
\]

From this we get:

\[
d_x = 3.516 \pm 0.0012~\text{\AA}
\]

\vspace{1em}

From COD Cu$_{1.78}$Se COD ID 1532676:

\[
a = 5.7464~\text{\AA}
\]

We transform the (hkl) = $(4~2~2)$  from FCC to hexagonal (HKL) using the transformation matrix:

\[
A =
\begin{pmatrix}
\frac{1}{2} & 0 & -\frac{1}{2} \\
-\frac{1}{2} & \frac{1}{2} & 0 \\
1 & 1 & 1
\end{pmatrix}
\]

\[
\begin{pmatrix}
H \\ K \\ L
\end{pmatrix}
=
A
\begin{pmatrix}
h \\ k \\ l
\end{pmatrix}
\quad \Rightarrow \quad
\begin{pmatrix}
H \\ K \\ L
\end{pmatrix}
=
\begin{pmatrix}
1 \\ -1 \\ 8
\end{pmatrix}
\]

\vspace{1em}

In our cubic samples, the out of plane orientation is the (111), and we are interested in the in plane componnt of our samples, which in heaxoganal coordinates is: 
\[
\begin{pmatrix}
H \\ K \\ L
\end{pmatrix}
=
\begin{pmatrix}
1 \\ -1 \\ 0
\end{pmatrix}
\]

\vspace{1em}

For the strain calculation, we are interested in comparing the $(1\bar{1}0)$ direction of our measurement vs Cu$_{1.78}$Se from COD ID 1532676

Using the inverse of the matrix \( A^{-1} \):

\[
A^{-1} =
\begin{pmatrix}
\frac{2}{3} & -\frac{2}{3} & \frac{1}{3} \\
\frac{2}{3} & \frac{1}{3} & -\frac{2}{3} \\
-\frac{1}{3} & \frac{2}{3} & \frac{2}{3}
\end{pmatrix}
\]

we get:

\[
\begin{pmatrix}
h \\ k \\ l
\end{pmatrix}
=
\begin{pmatrix}
\frac{2}{3} \\ -\frac{2}{3} \\ -\frac{2}{3}
\end{pmatrix}
\]

From Cu$_{1.78}$Se COD ID 1532676, the cubic lattice constant is:
\[
a = 5.7464~\text{\AA}
\]

The interplanar spacing is calculated as:

\[
d_{hkl} = \frac{a}{\sqrt{h^2 + k^2 + l^2}} = \frac{5.7464}{\sqrt{4}} = 3.514~\text{\AA}
\]

\[
{Strain_{cubic}} = 0.07\%
\]

\hrulefill
\vspace{10pt}
\section{RSM and Strain - rhombohedral phase}
\vspace{10pt}

\begin{figure}[H]  
    \centering
    \includegraphics[width=\linewidth]{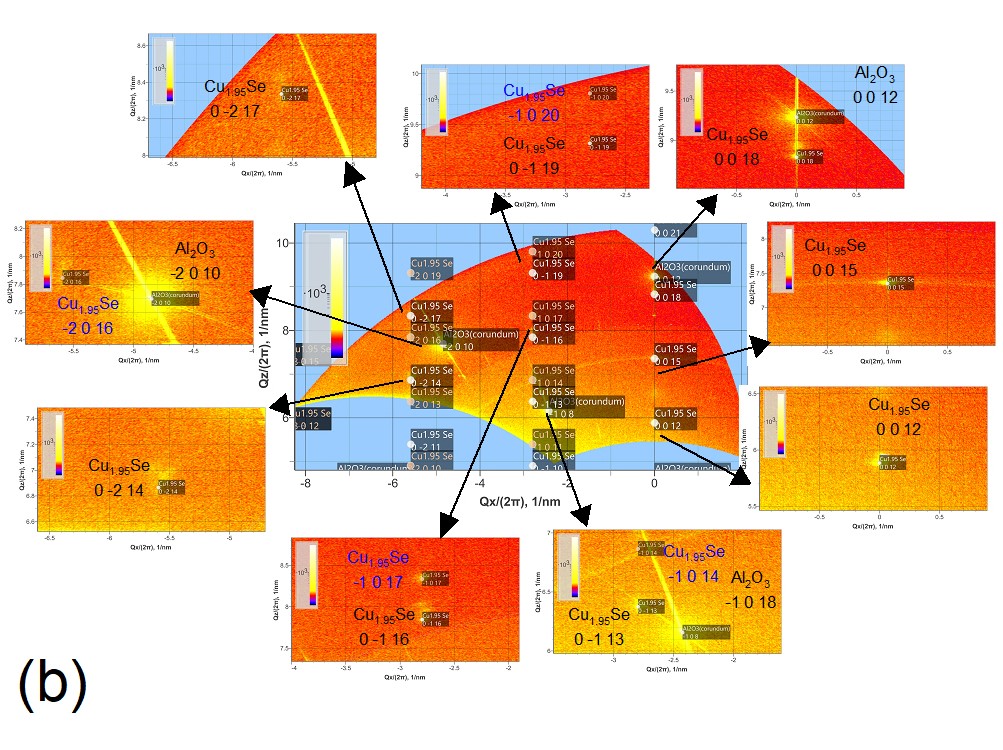}
    \caption{Detailed RSM comparison of Cubic phase.}
    \label{fig:cubic_phase}
\end{figure}

\textbf{Figure 2 shows the RSM for a rhombohedral phase sample. We will focuse on the (4 2 2) peak during this calculation and compare the results with Cu$_{1.95}$Se from COD ID 1556750. FROM RSM scan we have:}

\[
\frac{Q_{x(422)}}{2\pi} = -2.8439 \pm 0.00073~\text{nm}^{-1}
\]

From this we get:

\[
d_x = -3.580 \pm 0.0009~\text{\AA}
\]
Using this:

\[
\frac{1}{d^2} = \frac{4}{3} \cdot \frac{h^2 + hk + k^2}{a^2} + \frac{l^2}{c^2}
\]
We get: 
\[
a_{x, measured} = 4.134 \pm 0.001~\text{\AA}
\]
\vspace{1em}

From COD Cu$_{1.95}$Se COD ID 1556750:

\[
a = 4.1327~\text{\AA}
\]

\[
{Strain_{rhombohedral}} = 0.04\%
\]

\clearpage
\begin{figure}
    \centering
    \includegraphics[width=0.5\linewidth]{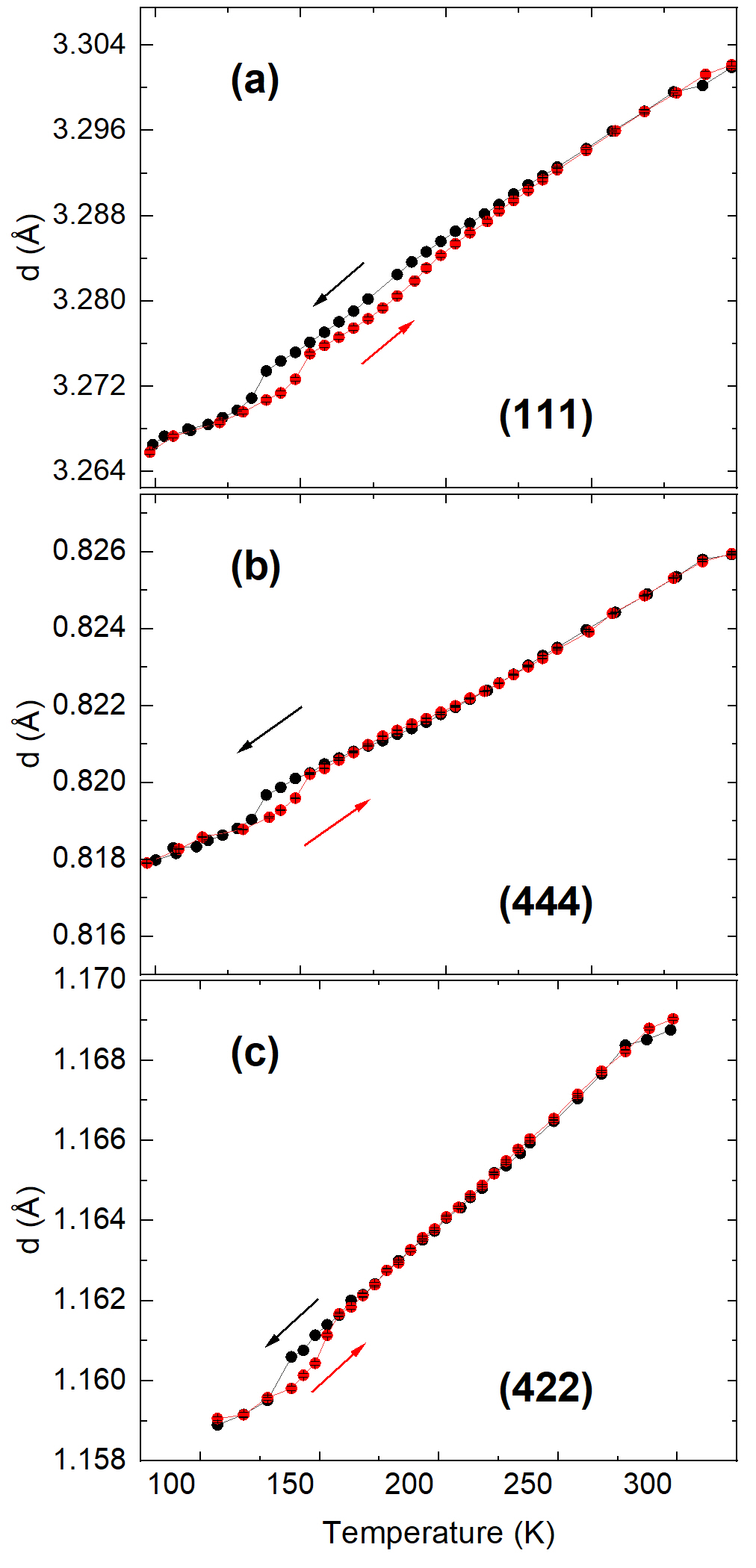}
    \caption{XRD vs. temperature: Comparison of $d$-spacing vs. temperature for the cubic phase—(a) (111), (b) (444) and (c) (422)
    \label{fig:Cubic XRD vs T S}}
\end{figure}

\begin{figure}
    \centering
    \includegraphics[width=\linewidth]{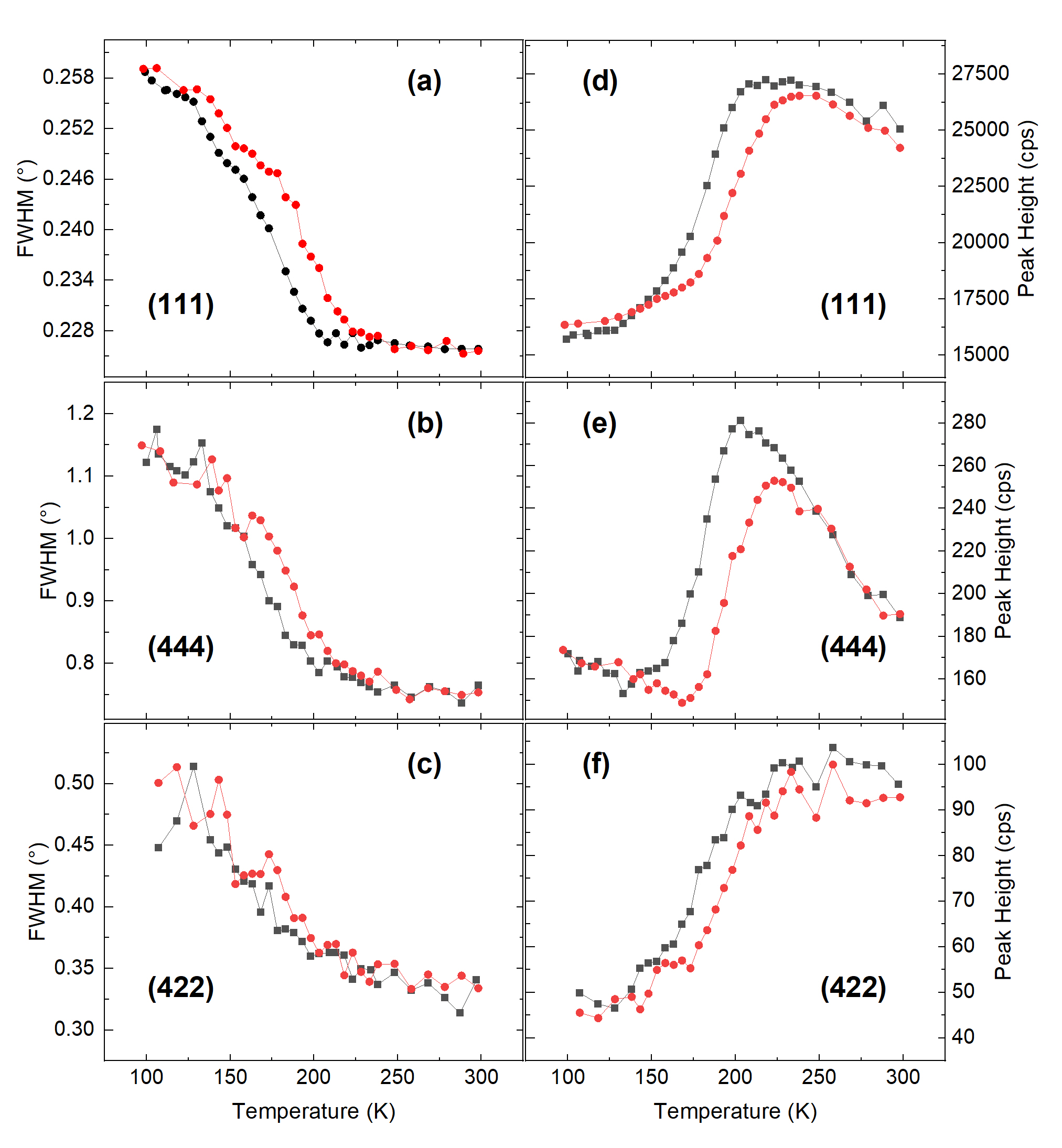}
    \caption{XRD vs. temperature: Comparison of $FWHM$ temperature for the cubic phase—(a) (111), (b) (444) and (c) (422); and the and Peak Height vs. temperature for the cubic phase—(d) (111), (e) (444) and (f) (422)
    \label{fig:Cubic XRD vs T S}}
\end{figure}

\begin{figure}
    \centering
    \includegraphics[width=\linewidth]{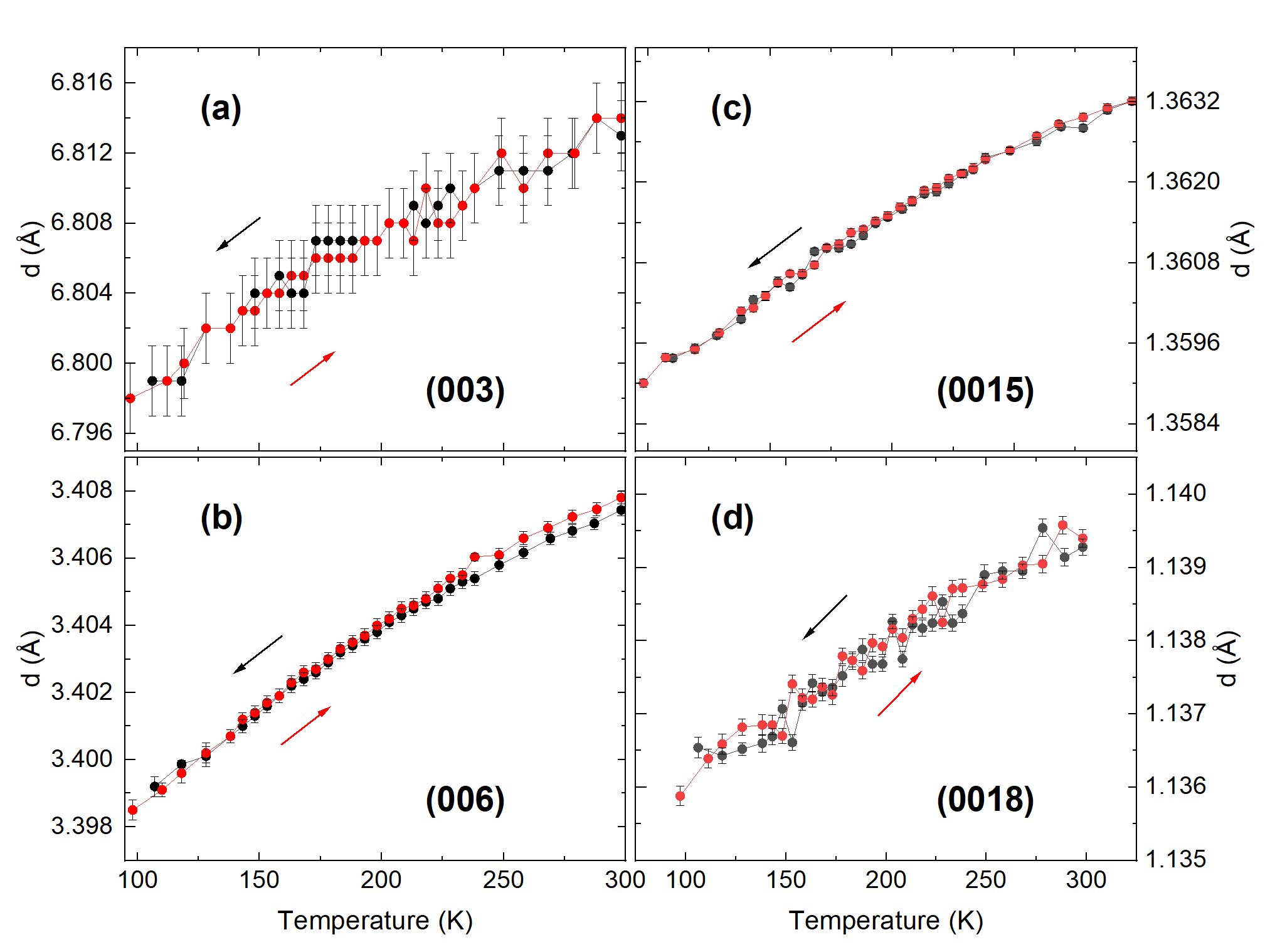}
    \caption{XRD vs. temperature: Comparison of $d$-spacing vs. temperature for the rhombohedral phase—(a) (003), (b) (006), (c) (0015) and (d) (0018)
    \label{fig:Rhombohedral XRD vs T S}}
\end{figure}

\begin{figure}
    \centering
    \includegraphics[width=0.5\linewidth]{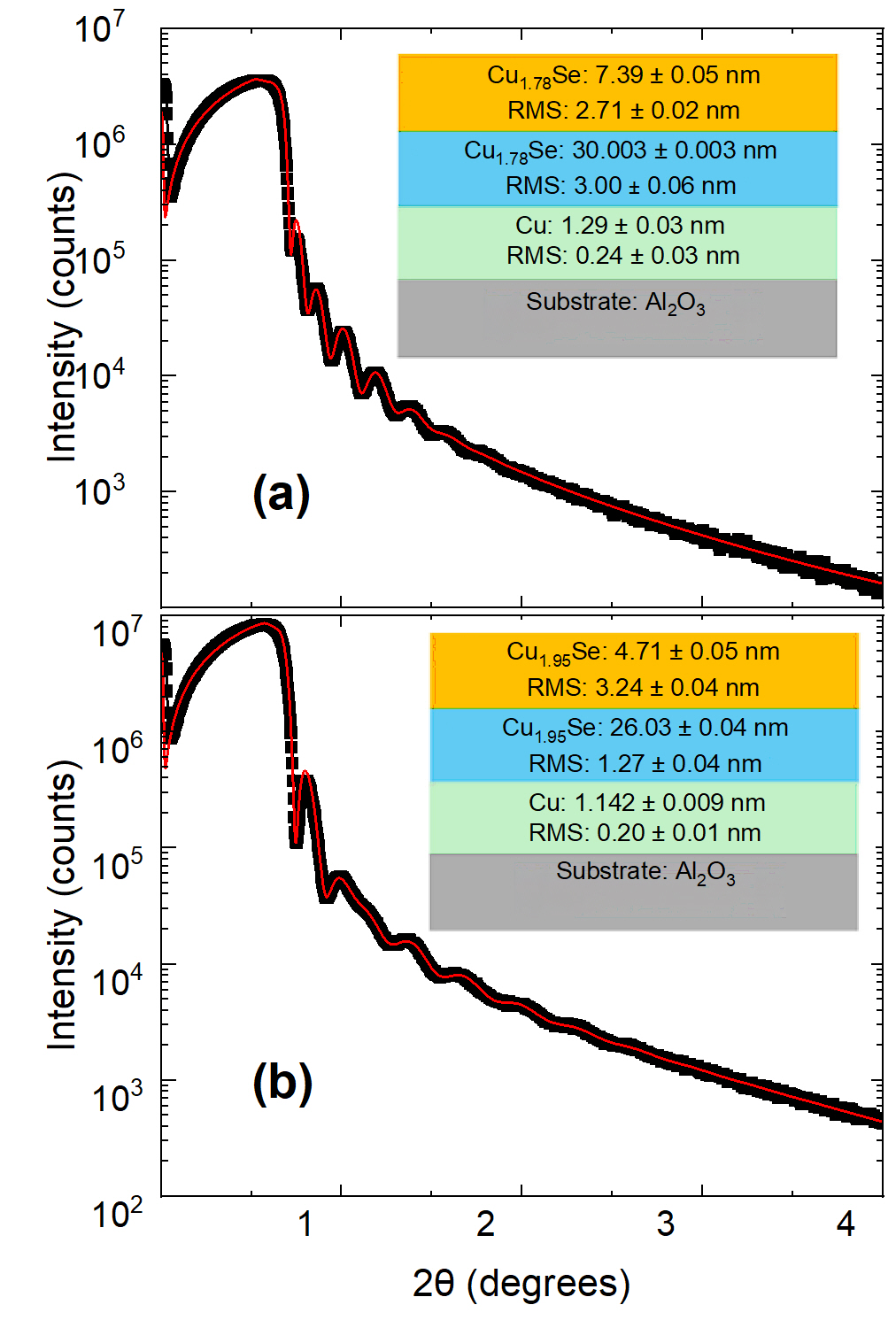}
    \caption{X-Ray Reflectivity (XRR) of the cubic phase (a) and the rhombohedral phase (b)
    \label{fig:XRD vs T S}}
\end{figure}

\begin{figure}
    \centering
    \includegraphics[width=0.5\linewidth]{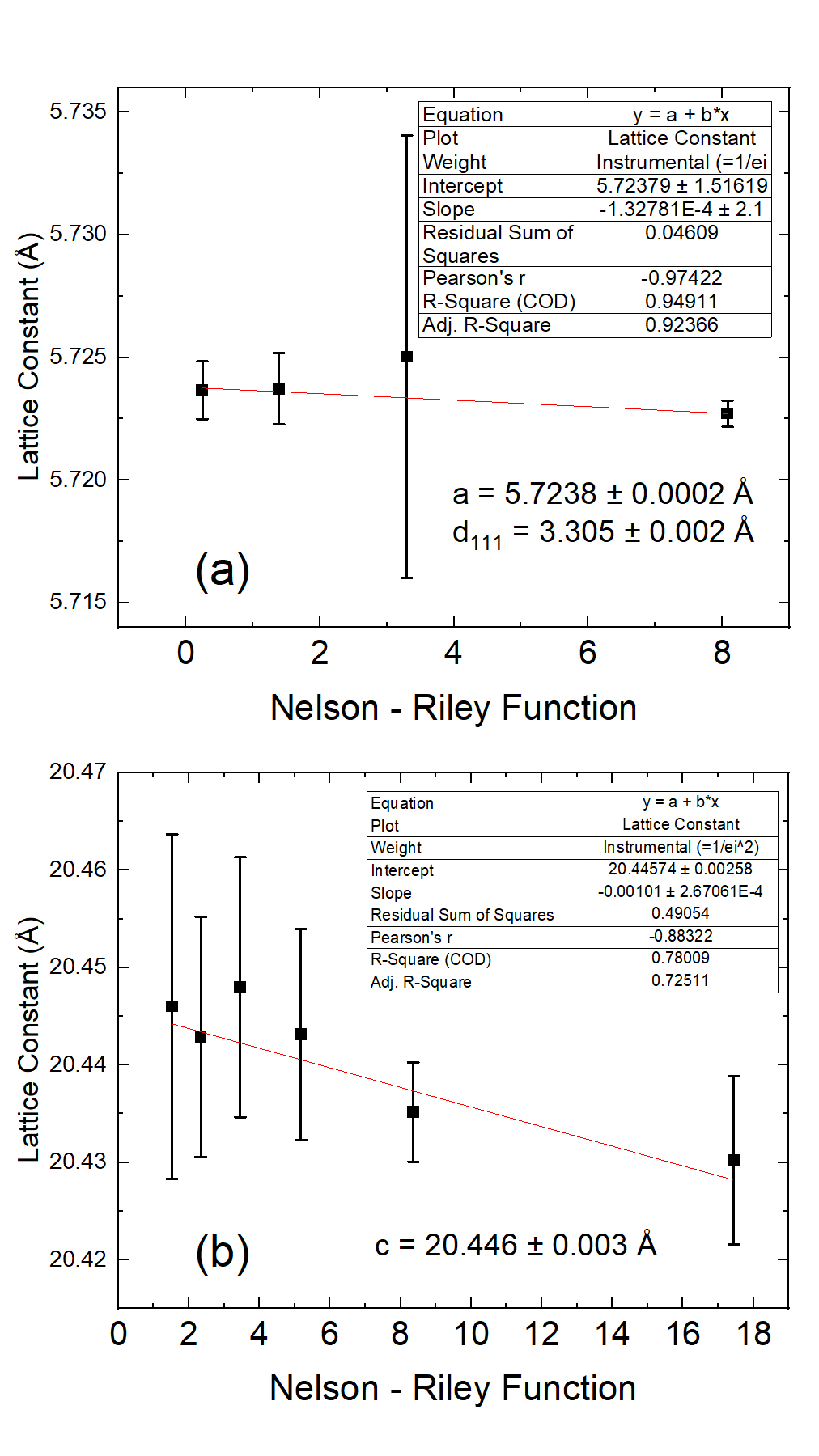}
    \caption{The lattice constant determined by plotting the calculated $a$ values as a function of the Nelson-Riley function and then fitting the resulting data to a straight line. (a) cubic phase (b) rhombohedral phase
    \label{fig:Lattice constant fit S}}
\end{figure}

\begin{figure}  
    \centering
    \includegraphics[width=\linewidth]{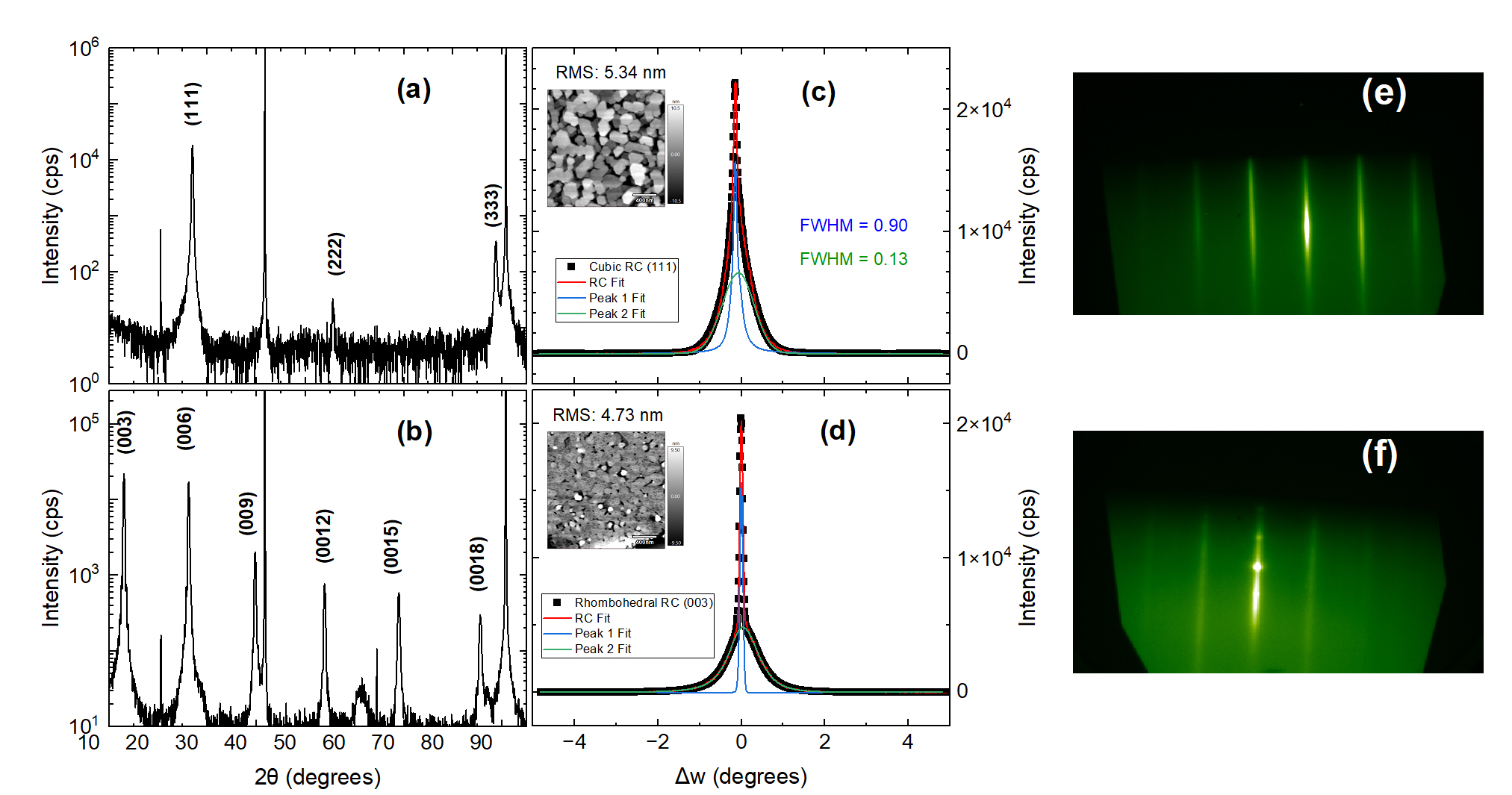}
    \caption{(Second cubic and rhombohedral samples) Comparison of structural characterization for two additional samples with the cubic and rhombohedral phases. (a) and (b) show XRD patterns with the scattering vector aligned along the growth direction for the cubic and rhombohedral phases, respectively. (c) and (d) display the corresponding rocking curves with AFM images shown as insets. (e) and (f) present RHEED patterns for the cubic and rhombohedral phases, respectively.
    \label{fig: Two additional samples RHEED_XRD_AFM S}}
\end{figure}

\begin{figure}
    \centering
    \includegraphics[width=0.5\linewidth]{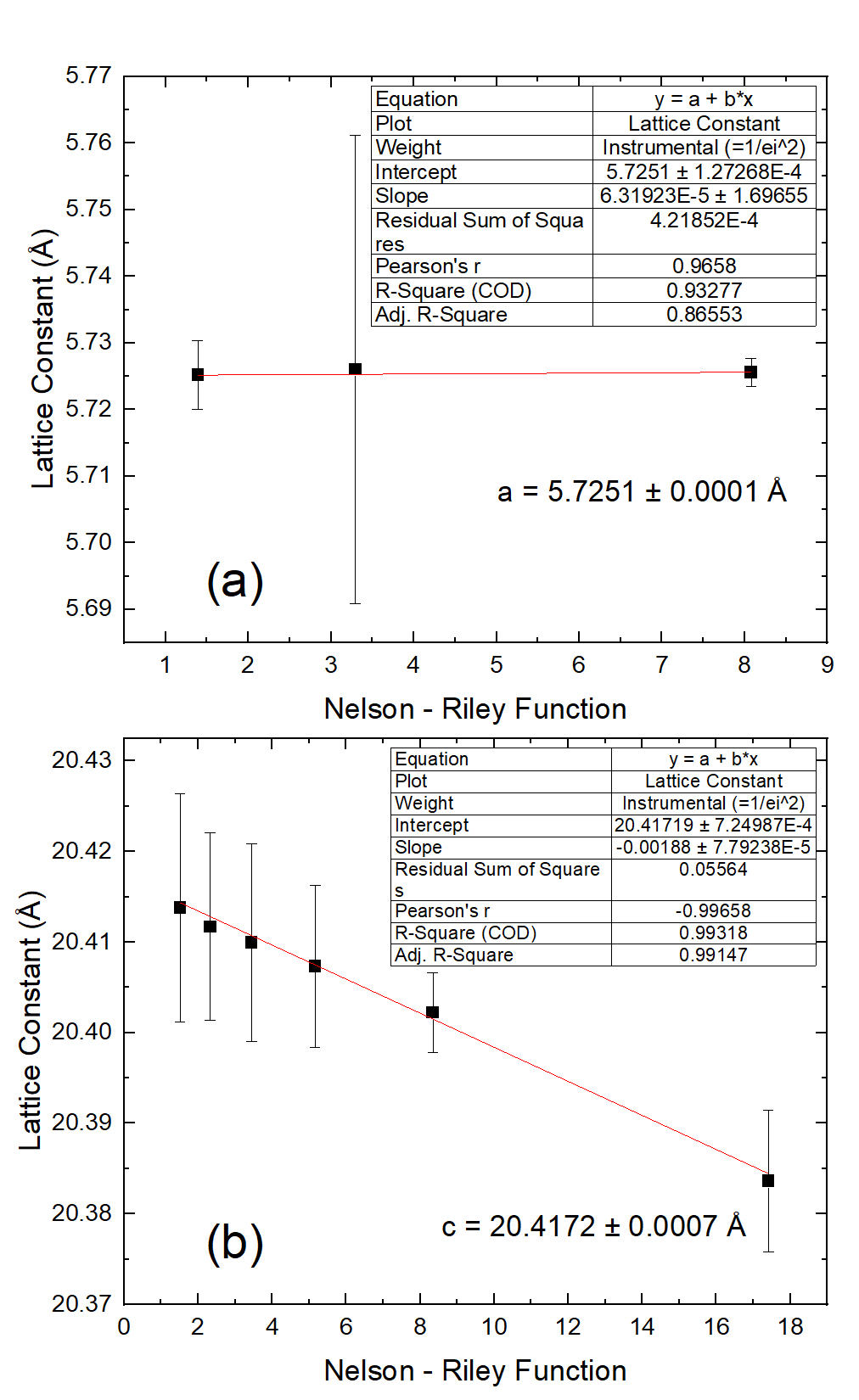}
    \caption{(Second cubic and rhombohedral samples) The lattice constant determined by plotting the calculated $a$ values as a function of the Nelson-Riley function and then fitting the resulting data to a straight line. (a) cubic phase (b) rhombohedral phase
    \label{fig:Two additional samples Lattice constant fit S}}
\end{figure}

\begin{figure}
    \centering
    \includegraphics[width=\linewidth]{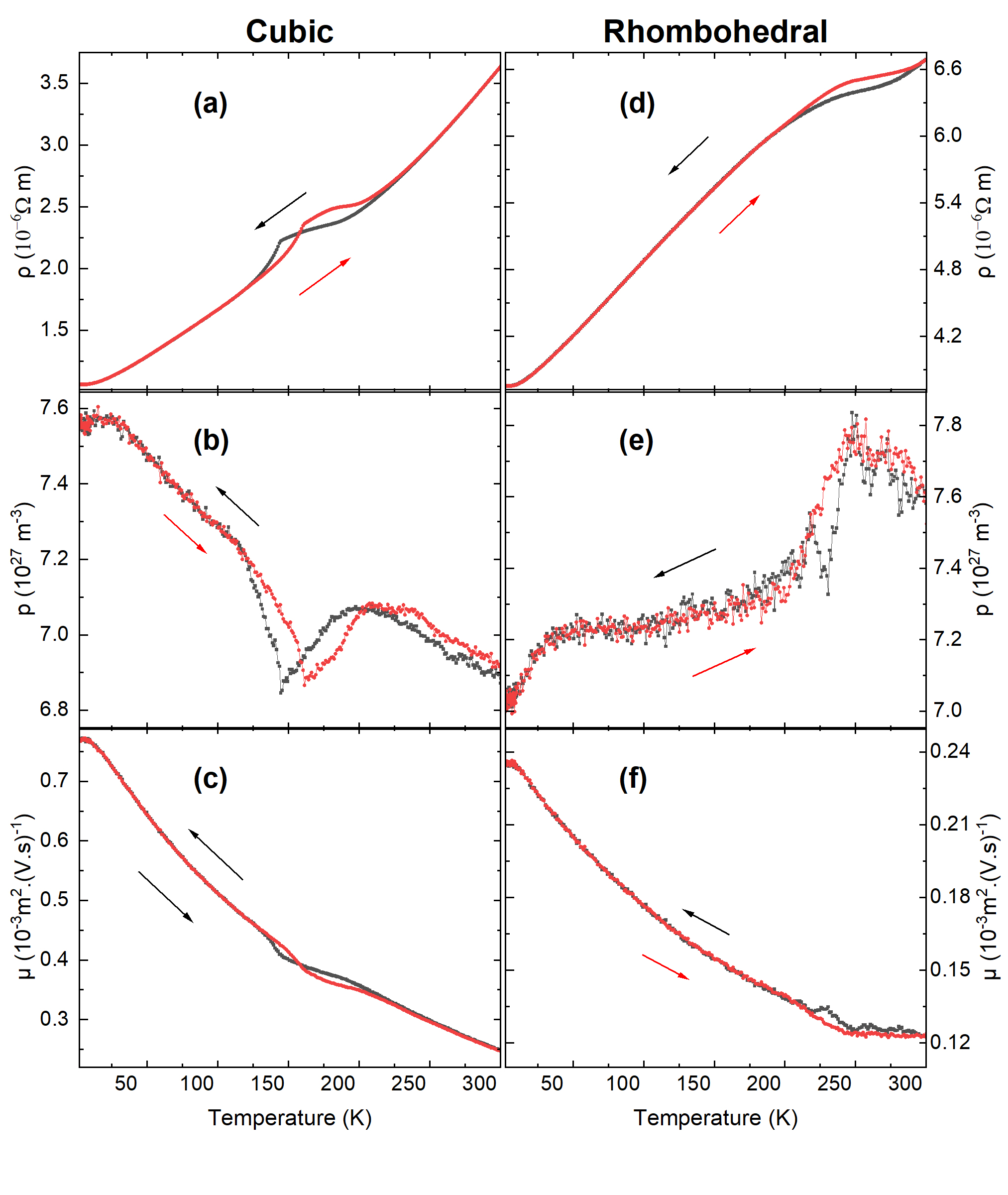}
    \caption{(Second cubic and rhombohedral samples) Resisitivity, carrier density, and mobility as functions of temperature for the second cubic [(a)-(c)] and the second rhombohedral [(d)-(f)] samples.
    \label{fig:Two additional samples's transport S}}
\end{figure}


\end{document}